\documentclass[aps,prb,twocolumn,showpacs,amsmath,tightenlines]{revtex4}
\usepackage{graphics}
\usepackage{bm}
\begin{document}

\title{Conductance of a quantum wire at low electron density}

\author{K. A. Matveev} 

\affiliation{Argonne National Laboratory, 9700 S Cass Ave., Argonne, IL
  60439, USA} 

\affiliation{Department of Physics, Duke University, Durham, NC 27708,
  USA}

\date{May 23, 2004} 

\begin{abstract}
  We study the transport of electrons through a long quantum wire
  connecting two bulk leads.  As the electron density in the wire is
  lowered, the Coulomb interactions lead to short-range crystalline
  ordering of electrons.  In this Wigner crystal state the spins of
  electrons form an antiferromagnetic Heisenberg spin chain with
  exponentially small exchange coupling $J$.  Inhomogeneity of the
  electron density due to the coupling of the wire to the leads results in
  violation of spin-charge separation in the device.  As a result the
  spins affect the conductance of the wire.  At zero temperature the
  low-energy spin excitations propagate freely through the wire, and its
  conductance remains $2e^2/h$.  Since the energy of the elementary
  excitations in the spin chain (spinons) cannot exceed $\pi J/2$, the
  conductance of the wire acquires an exponentially small negative
  correction $\delta G \propto - \exp(-\pi J/2T)$ at low temperatures
  $T\ll J$.  At higher temperatures, $T\gg J$, most of the spin
  excitations in the leads are reflected by the wire, and the conductance
  levels off at a new universal value $e^2/h$.
\end{abstract}


\maketitle

\section{Introduction}

The quantization of conductance of one-dimensional (1D) electron systems
in units of $2e^2/h$ was first observed in experiments with quantum point
contacts.\cite{wees,wharam} The latter consist of a short (well under 1
$\mu$m) 1D constriction connecting two bulk two-dimensional leads.
Further progress in fabrication of low-disorder devices resulted in
observation\cite{tarucha,yacoby} of similar conductance quantization in
quantum wires of several microns in length.  Experimentally the
quantization is observed as very flat plateaus in the dependence of linear
conductance on the voltage at the gate controlling the electron density in
the wire.

In a number of recent
experiments\cite{thomas,thomas1,cronenwett,kristensen1,kristensen,bkane,thomas2,reilly1,reilly2}
deviations of conductance from perfect quantization have been observed at
low electron density.  These deviations manifest themselves as negative
corrections to the conductance at the beginning of the first quantized
plateau.  The correction is usually small at the lowest temperatures
available, but becomes significant at $T\sim 1$K.  In typical
samples\cite{thomas,thomas1,cronenwett} the conductance levels off at high
temperatures and forms a quasi-plateau at about $0.7\times (2e^2/h)$.
This phenomenon is often referred to as the {\em 0.7 structure}.  Despite
the numerous theoretical
attempts\cite{wang,bruus1,reimann,spivak,flambaum,rejec,bruus,hirose,sushkov,tokura,meir,seelig,starikov}
at the interpretation of the 0.7 structure, its origin remains unclear.

The analysis of the existing experimental data shows that the 0.7
structure is sensitive to the length of the one-dimensional region
connecting the two-dimensional leads.  The structure tends to be
relatively weak in the short contacts,\cite{kristensen1,kristensen} where
no quasi-plateau is observed even at high temperatures.  On the other
hand, in longer samples\cite{bkane,thomas2,reilly1,reilly2} the plateau is
observed even at the lowest temperatures available.  The effect is also
somewhat stronger, with the quasi-plateau moving to a lower value of
conductance $G\approx0.5\times (2e^2/h)$.

In this paper we consider conductance of a long quantum wire in the regime
of low electron density $n$.  Focusing on the case of a GaAs device, we
assume quadratic energy spectrum for free electrons $\epsilon(p)=p^2/2m$,
where $m$ is the effective mass of the electrons.  The typical kinetic
energy of an electron at zero temperature is of the order of the Fermi
energy, $E_F=(\pi\hbar n)^2/8m$.  It is important to note that at low
density $n\ll a_B^{-1}$ the kinetic energy is small compared to the
typical energy $e^2n/\varepsilon$ of Coulomb interaction between
electrons, where $\varepsilon$ is the dielectric constant, and
$a_B=\varepsilon\hbar^2/me^2$ is the effective Bohr radius of the
material.  Thus in the limit of low density the electrons can be viewed as
classical particles placed at equidistant positions to minimize the
Coulomb repulsion.  Such a picture was first proposed by
Wigner\cite{wigner} and will be referred to as the {\em Wigner crystal}.

Although the quantum fluctuations of electrons near their equilibrium
positions destroy the long-range order in the Wigner crystal, its
short-range structure strongly affects the transport through the wire.  In
particular, the electrons occupying well-defined sites of a Wigner lattice
can be viewed as an antiferromagnetic spin chain with exponentially small
exchange constant $J$.  The appearance of a new energy scale $J\ll E_F$
significantly affects the physics of the electronic transport through the
wire.  This effect is most important at intermediate temperatures, $J\ll T
\ll E_F$, where it results in a considerable suppression of the
conductance of the wire.

The physics of this phenomenon is controlled by the effect of spin-charge
separation in one-dimensional interacting electron systems.  The latter
refers to the fact that fermionic quasiparticles cannot be viewed as
elementary excitations of the system, i.e., it no longer behaves as a
Fermi liquid.  Instead, the system displays Luttinger liquid behavior, and
the elementary excitations are bosonic waves of charge and spin densities
propagating at different velocities.  As a result, when an electron enters
the one-dimensional region from one of the leads, it is decomposed into
charge and spin waves.  At low temperature $T\ll J$ both waves pass
through the wire, and upon reaching the other lead they reassemble into a
fermionic quasiparticle.  This process can be interpreted as perfect
transmission of electrons through the wire and gives the standard value of
conductance $G=2e^2/h$.  On the other hand, at $T\gg J$, the bandwidth
$\sim J$ of the spin excitations in the wire is small compared to their
typical energy $\sim T$.  As a result, only the charge excitations pass
through the wire, whereas the spin ones are reflected back to the lead.
We show below that this additional scattering of spin excitations by the
wire reduces its conductance to $e^2/h$

In Sec.~\ref{sec:spin_charge} we study the applicability of the Luttinger
liquid description to the 1D Wigner crystal and show that it is only valid
at energy scales below $J$.  On the other hand, the property of
spin-charge separation is more general and persists at energy scales above
$J$.  We review the known results for the conductance of a quantum wire in
Sec.~\ref{sec:separation}.  The most important consequence of the
spin-charge separation in quantum wires is the conclusion that spin
degrees of freedom do not affect the conductance, which remains quantized
at $2e^2/h$.  In Sec.~\ref{sec:nonseparation} we show that the spin-charge
separation is violated when the wire is connected to the leads.  As a
result the spin subsystem affects the propagation of electric charge
through the wire and contributes to its resistance.  This contribution is
studied in Sec.~\ref{sec:conductance}, where we find that at $J\ll T$ the
conductance of the device reduces from $2e^2/h$ to $e^2/h$.  The relation
of our results to experimental measurements of conductance of quantum
wires is discusses in Sec.~\ref{sec:discussion}.  A brief summary of some
of our results has been reported in Ref.~\onlinecite{PRL04}.

\section{Spin-charge separation in quantum wires}
\label{sec:spin_charge}

It has been known since the seventies\cite{dzyaloshinskii} that the
low-energy excitations of a 1D system of interacting electrons are the
charge and spin waves propagating independently of each other at different
velocities.  This result is valid at the energy scales low compared to the
bandwidths $D_\rho$ and $D_\sigma$ of the charge and spin excitations.  In
the case of not very strong interactions both bandwidths are of the order
of the Fermi energy, and the picture of completely separated charge and
spin excitations is appropriate at $T\ll E_F$.

At low electron density $n\ll a_B^{-1}$ the interactions between electrons
are strong.  We show below that as a result the velocity of the spin
excitations is greatly reduced, and their bandwidth $D_\sigma\sim J$
becomes much smaller than $E_F$.  A similar effect is known to occur in
the strong interaction limit of some lattice models, such as the Hubbard
model.\cite{coll} In this regime, the description of the spin excitations
in the language of non-interacting spin waves is applicable only at very
low temperatures $T\ll J$.  We show in this section that a generalized
picture of decoupled charge and spin excitations remains valid even at
$T\gtrsim J$.  In this picture the charge excitations are still given by
the waves of charge density (plasmons), and the spin waves are replaced
with the excitations of a Heisenberg spin chain.

\subsection{Luttinger-liquid picture of one-dimensional electrons systems}

The problems involving low-energy properties of interacting 1D electron
systems are conveniently described in the framework of the bosonization
technique.\cite{haldane,schulz} The first step in this approach is to
linearize the spectrum of electrons near the Fermi level, thereby
replacing the quadratic dispersion law $\epsilon(k)=\hbar^2 k^2/2m$ with
the linear one.  In this {\em Tomonaga-Luttinger model\/} the electrons
are separated in two branches, the left- and right-movers, with energies
$\epsilon_{L,R}(k)=\hbar v_F(\mp k - k_F)$, where $v_F$ and $k_F$ are the
Fermi velocity and Fermi wavevector.  One can then present the fermionic
field operators $\psi_{L,\lambda}$ and $\psi_{R,\lambda}$ in terms of
fields $\phi_\lambda$ and $\theta_\lambda$ satisfying bosonic commutation
relations $[\phi_\lambda(x), \partial_y\theta_{\lambda'}(y)] =
i\pi\delta(x-y)\delta_{\lambda\lambda'}$ using the following rule
\begin{subequations}
\begin{eqnarray}
  \label{eq:bosonization}
  \psi_{L,\lambda}(x) &=& \frac{\eta_{L,\lambda}}{\sqrt{2\pi\alpha}} 
                    e^{-ik_Fx} e^{i\phi_\lambda(x)-i\theta_\lambda(x)},
  \\
  \psi_{R,\lambda}(x) &=& \frac{\eta_{R,\lambda}}{\sqrt{2\pi\alpha}} 
                    e^{ik_Fx} e^{-i\phi_\lambda(x)-i\theta_\lambda(x)}.
\end{eqnarray}
\end{subequations}
Here $\lambda=\uparrow,\downarrow$ is the spin index, $\alpha$ is the
short distance cutoff, $\eta_{L,\lambda}$ and $\eta_{L,\lambda}$ are
Majorana fermion operators.\cite{schulz}

In terms of the bosonic variables the Hamiltonian of an interacting
1D electron system takes the form\cite{schulz,giamarchi}
\begin{equation}
  \label{eq:separation}
  H = H_\rho + H_\sigma,
\end{equation}
where the two terms $H_\rho$ and $H_\sigma$ describe the excitations of
the charge and spin degrees of freedom, respectively, and have the forms
\begin{eqnarray}
  \label{eq:H_rho}
  H_\rho&=&\int\frac{\hbar u_\rho}{2\pi}\left[\pi^2 K_\rho \Pi_\rho^2 
                         +K_\rho^{-1}(\partial_x\phi_\rho)^2\right]dx,
\\
  \label{eq:H_sigma}
  H_\sigma&=&\int\frac{\hbar u_\sigma}{2\pi}
                         \left[\pi^2 K_\sigma \Pi_\sigma^2 
                         +K_\sigma^{-1}(\partial_x\phi_\sigma)^2\right]dx
\nonumber\\
          &&+\frac{2g_{1\perp}}{(2\pi\alpha)^2}\int 
             \cos\left[\sqrt8 \phi_\sigma(x)\right] dx.
\end{eqnarray}
Here the new bosonic fields $\phi_{\rho,\sigma} = (\phi_\uparrow \pm
\phi_\downarrow)/\sqrt2$ and $\Pi_{\rho,\sigma} =
\partial_x(\theta_\uparrow \pm \theta_\downarrow)/\pi\sqrt2$ satisfy the
standard commutation relations $[\phi_\alpha(x), \Pi_{\alpha'}(y)] =
i\delta(x-y)\delta_{\alpha\alpha'}$ and represent the excitations of
the charge and spin degrees of freedom.  The Hamiltonian
(\ref{eq:separation})--(\ref{eq:H_sigma}) depends on five parameters
determined by the interactions between electrons: velocities $u_\rho$ and
$u_\sigma$ of the charge and spin excitations, dimensionless parameters
$K_\rho$ and $K_\sigma$, and the matrix element $g_{1\perp}$ of spin-flip
scattering of a left-moving electron and a right-moving one.  In the
absence of interactions, $u_\rho=u_\sigma=v_F$, $K_\rho=K_\sigma=1$, and
$g_{1\perp}=0$.  The bosonized Hamiltonian correctly describes the charge
and spin excitations of the system with energies below the respective
bandwidths $D_{\rho,\sigma} \sim \hbar n u_{\rho,\sigma}$. 

In the most interesting case of repulsive interactions $K_\rho<1$.  The
coupling constant $g_{1\perp}$ is positive and scales to zero at low
energy scales $D$,
\begin{equation}
  \label{eq:marginal}
  g_{1\perp} = \frac{g_{1\perp}}
                    {1+\frac{g_{1\perp}}{\pi
                    u_\sigma}\ln\frac{D_\sigma}{D}}. 
\end{equation}
The parameter $K_\sigma$ renormalizes along with
$g_{1\perp}$, approaching the value $K_\sigma=1$ required by the SU(2)
symmetry as $K_\sigma=1+g_{1\perp}/2\pi u_\sigma$,
Ref.~\onlinecite{schulz,giamarchi}.

Since the sine-Gordon term in Eq.~(\ref{eq:H_sigma}) vanishes at
$D/D_\sigma\to 0$, the Hamiltonian (\ref{eq:separation}) becomes quadratic
and describes a Luttinger liquid.\cite{haldane} The latter represents a
stable fixed point of the problem, so the description of the system based
upon the Hamiltonian (\ref{eq:separation})--(\ref{eq:H_sigma}) is expected
to be valid in a broad range of interaction strengths.  It is not
immediately obvious, however, that the above picture is applicable at low
electron density $n$, i.e., when the interactions are so strong that the
electrons form a Wigner crystal.  Indeed, a Fourier expansion of the
wavefunction of an electron localized in a small region of size $a\ll
n^{-1}$ near a given lattice site involves the wavevectors in a broad
range $\delta k\sim 1/a \gg n\sim k_F$.  Thus the standard procedure of
linearization of the electronic spectrum near $k_F$ leading to the
Tomonaga-Luttinger model is not justified in this case.  In addition, each
electron is constructed out of waves with both positive and negative
wavevectors, and the picture of two separate branches of left- and
right-moving particles is not applicable to the Wigner crystal.  We show
now that even though the conventional derivation leading to
Eqs.~(\ref{eq:separation})--(\ref{eq:H_sigma}) is not justified, this
Hamiltonian does describe the low-energy properties of a 1D Wigner
crystal.

\subsection{Charge and spin excitations in a Wigner crystal}
\label{sec:wigner}

At low electron density, $na_B\ll1$, the properties of the system are
dominated by the Coulomb repulsion, and the electrons occupy fixed
positions on the Wigner lattice.  The first correction to this picture is
due to the small vibrations of the lattice, analogous to phonons in
conventional crystals.  In the long-wavelength limit these phonons can be
described in the framework of elasticity theory.  In this approach the
crystal is viewed as an elastic medium.  Its motion is described in terms
of the displacement $u(x)$ of the medium at point $x$ from its equilibrium
position and the momentum density $p(x)$.  The energy of the system can
then be written as a sum of kinetic and potential energies,
\begin{equation}
  \label{eq:Wigner_Hamiltonian}
  H = \int \left[\frac{p^2}{2mn} + \frac12 mns^2 (\partial_x u)^2\right]dx.
\end{equation}
The second term here is $\frac12(\partial^2 E/\partial n^2)(\delta n)^2$,
where $E$ is the energy of the resting medium per unit length, and the
density perturbation $\delta n$ is proportional to the deformation of the
medium, $\delta n = -n\,\partial_x u$.  The parameter
$s=\sqrt{(n/m)(\partial^2 E/\partial n^2)}$ has the meaning of the speed
of density waves (plasmons) in the Wigner crystal.

The speed of plasmons in a 1D system with true Coulomb interactions
between electrons diverges in the limit of long wavelength.  In practice,
however, the interactions between electrons are usually screened at large
distances by a remote metal gate.  In the model where the gate is a
conducting plane at a large distance $d\gg n^{-1}$ from the Wigner
crystal, the speed of plasmons is
\begin{equation}
  \label{eq:plasmon_speed}
  s=\sqrt{\frac{2e^2n}{\varepsilon m}\ln(\zeta nd)},
\end{equation}
where $\zeta\approx 8.0$, Ref.~\onlinecite{ruzin}. 

The classical Hamilton function (\ref{eq:Wigner_Hamiltonian}) can be
quantized by imposing commutation relations
$[u(x),p(y)]=i\hbar\delta(x-y)$.  The resulting Hamiltonian describes the
propagation of the electron density excitations in a Wigner crystal, and
is completely analogous to the term $H_\rho$ in the Hamiltonian
(\ref{eq:separation})--(\ref{eq:H_sigma}) of the Luttinger liquid.
Comparing the commutation relations of the bosonic fields entering
Hamiltonians (\ref{eq:H_rho}) and (\ref{eq:Wigner_Hamiltonian}), and taking
into account the expressions for the density perturbation $\delta n =
-\frac{\sqrt2}{\pi} \partial_x\phi_\rho = -n\,\partial_x u$, we identify
the fields as
\begin{equation}
  \label{eq:field_identities}
  u(x) = \frac{\sqrt2}{\pi n}\, \phi_\rho(x), 
  \quad 
  p(x) = \frac{\pi n\hbar}{\sqrt2}\, \Pi_\rho(x).
\end{equation}
Using these expressions we can relate the parameters in the Hamiltonian
$H_\rho$ to the properties of the Wigner crystal as follows
\begin{equation}
  \label{eq:parameter_relations}
  u_\rho = s, \quad K_\rho = \frac{v_F}{s}.
\end{equation}
Here $v_F = \pi\hbar n/2m$ is the Fermi velocity in a non-interacting
Fermi gas of density $n$.

The electrons in a Wigner crystal are repelled from each other by strong
Coulomb forces.  In the harmonic chain approximation we used so far the
electrons are allowed to move about their equilibrium positions; however,
the amplitude of these oscillations remains small.  As a result the
electrons never move from one site on the Wigner lattice to another, and
can be viewed as distinguishable particles.  Therefore the energy of the
Wigner crystal state in this approximation does not depend on the electron
spins.

To account for the spin dependence, one has to include the processes in
which electrons tunnel through the Coulomb potential repelling them.
Considering a pair of electrons at two neighboring sites of the Wigner
lattice, one notices that depending on their total spin, the two electrons
occupy either a symmetric or an antisymmetric state in the respective
double-well potential.  Thus the energy of the pair contains a term $J\,
{\bm S}_{1}\cdot{\bm S}_{2}$, where $J>0$ is the difference of energies of
the antisymmetric and symmetric states, and ${\bm S}_{i}$ are the
operators of electron spins.\cite{hausler1} Taking into account all the
nearest neighbor sites, we find that the spin properties of the Wigner
crystal are described by the Hamiltonian of an antiferromagnetic
Heisenberg spin chain
\begin{equation}
  \label{eq:spin-chain}
  H_\sigma = \sum_l J\, {\bm S}_{l}\cdot{\bm S}_{l+1}.
\end{equation}
Since the exchange is due to tunneling, the constant $J$ is exponentially
small,
\begin{equation}
  \label{eq:J}
  J = J^* \exp\left(-\frac{\eta}{\sqrt{na_B}}\right).
\end{equation}
To accurately evaluate $J$ one has to take into account the fact that when
two neighboring electrons tunnel through the Coulomb barrier separating
them, all other electrons also move.  H\"ausler\cite{hausler} suggested an
approximation that neglects the motion of other electrons; his result
corresponds to the value of $\eta\approx 2.87$ in a finite chain of 15
electrons.  We solve this model in the case of an infinite chain  in
Appendix~\ref{sec:WKB} and obtain the value of $\eta\approx 2.82$.  We also
estimate the prefactor as
\begin{equation}
  \label{eq:prefactor}
  J^*\approx 1.79\, \frac{E_F}{(n a_B)^{3/4}},
\end{equation}
where $E_F=(\pi\hbar n)^2/8m$ is the Fermi energy of a non-interacting
electron gas of density $n$.

The spin part (\ref{eq:spin-chain}) of the Hamiltonian of a 1D Wigner
crystal is very different from that of a weakly interacting electron gas,
Eq.~(\ref{eq:H_sigma}).  It is easy to show, however, that at low energies
$D\ll J$ the two Hamiltonians are equivalent.  To accomplish that we use
the standard procedure\cite{schulz,giamarchi} of bosonization of
spin chains.  The first step is to perform the Jordan-Wigner
transformation
\begin{equation}
  \label{eq:Jordan-Wigner}
  S_l^z = a_l^\dagger a_l - \frac12,
  \quad
  S_l^x + iS_l^y 
      = a_l^\dagger\exp\left(i\pi\sum_{j=1}^{l-1} a_j^\dagger a_j\right),
\end{equation}
which expresses the spin operators in terms of creation and destruction
operators $a^\dagger$ and $a$ of spinless fermions.  In terms of these
operators the Hamiltonian (\ref{eq:spin-chain}) becomes 
\begin{eqnarray}
    H_\sigma &=& \frac12 \sum_l J\biggl[
               \left(a_l^\dagger a_{l+1} + a_{l+1}^\dagger a_{l}\right)
\nonumber\\
             && + 2\left(a_l^\dagger a_l + \frac12\right)
                       \left(a_{l+1}^\dagger a_{l+1} + \frac12\right)
                 \biggl].
  \label{eq:fermionic}
\end{eqnarray}
Thus the Heisenberg spin chain (\ref{eq:spin-chain}) is equivalent to the
model (\ref{eq:fermionic}) of interacting lattice fermions.

The second step is to bosonize the Hamiltonian (\ref{eq:fermionic}).  At
low energies one can replace the lattice model (\ref{eq:fermionic}) with a
continuous one, $a_l\to a(y)$, linearize the spectrum of the fermions near
the Fermi level, and then apply a bosonization transformation
\begin{equation}
  \label{eq:bosonization_spin-chain}
  a_{L,R}(y)=\frac{\eta_{L,R}}{\sqrt{2\pi\alpha}} 
             e^{\mp ik_Fy}
             e^{\pm i\phi_\sigma(y)/\sqrt{2}-i\sqrt{2}\,\theta_\sigma(y)}.
\end{equation}
The resulting bosonized Hamiltonian of the spin chain is
equivalent\footnote{The numerical coefficients in the Hamiltonian
  (\ref{eq:H_sigma}) are different from those obtained by conventional
  bosonization\cite{schulz,giamarchi} of the Heisenberg model.  The
  discrepancy is due to the fact that the standard procedure uses
  different bosonic fields $\phi=\phi_\sigma/\sqrt{2}$ and
  $\theta=\sqrt{2}\theta_\sigma$.} to Eq.~(\ref{eq:H_sigma}).  The value
of the speed $u_\sigma$ of the spin excitations is easily deduced from the
Bethe ansatz solution\cite{des_cloizeaux,faddeev} of the Heisenberg model,
\begin{equation}
  \label{eq:u_sigma}
  u_\sigma = \frac{\pi J}{2\hbar n}.
\end{equation}

Thus we have established that the bosonized Hamiltonian
(\ref{eq:separation})--(\ref{eq:H_sigma}) adequately describes the
low-energy properties of not only weakly interacting electron systems, but
also of a 1D Wigner crystal state at $na_B\ll 1$.  However, it is
important to keep in mind that the applicability of the bosonized
description to the Wigner crystal is limited to very low temperatures
$T\ll J$.  Given the exponential dependence (\ref{eq:J}) of the exchange
constant on density, this condition can be easily violated even at fairly
low temperatures.  In this case one has to use the more complicated form
(\ref{eq:spin-chain}) of the Hamiltonian $H_\sigma$.  We show in
Sec.~\ref{sec:conductance} that this breakdown of the Luttinger liquid
picture gives rise to significant deviations of conductance of quantum
wires from the quantized value $2e^2/h$.

\subsection{Spin-charge separation at ultralow electron densities}

The Wigner crystal picture discussed in Sec.~\ref{sec:wigner} relies on
the long-range nature of the Coulomb interaction potential
$V(x)=e^2/\varepsilon|x|$.  In general, a 1D electron system forms a
Wigner crystal state at $n\to 0$ only if the interaction potential decays
slower than $1/|x|^2$ at $x\to\infty$.  Indeed, for potential $V(x)\propto
1/|x|^\gamma$ the interaction energy of two electrons at the typical
interparticle distance $n^{-1}$ is $V\propto n^\gamma$, whereas the
kinetic energy $E_F\propto n^2$.  Thus at $\gamma>2$ the interaction
energy is negligible at $n\to 0$.

The electron density in quantum wires is usually controlled by applying
voltage to metal gates.  The presence of a gate affects the
electron-electron interactions at large distances.  For instance, if the
gate is modeled by a conducting plane at a distance $d$ from the wire,
the interactions between electrons become
\begin{equation}
  \label{eq:screened_interaction}
  V(x)=\frac{e^2}{\varepsilon}
       \left(\frac{1}{|x|}-\frac{1}{\sqrt{x^2+(2d)^2}}\right).
\end{equation}
The screening of the Coulomb potential by the gate reduces the potential
(\ref{eq:screened_interaction}) to $V(x)=2e^2d^2/\varepsilon |x|^3$ at
large $|x|$.  Therefore, the Wigner crystal picture fails in the limit
$n\to0$.  Comparing the interaction potential at the interparticle
distance $V(n^{-1})$ with the Fermi energy of electrons, one concludes
that within the model (\ref{eq:screened_interaction}) the Wigner crystal
state exists only in the density range $a_B/d^2\ll n\ll a_B^{-1}$.

As long as the system is in the Wigner crystal state, its spin excitations
are described by the  Heisenberg model (\ref{eq:spin-chain}).  However,
the expression (\ref{eq:J}) for the coupling constant $J$ relies on the
pure Coulomb interaction between electrons.  In the case of interaction
potential screened by the gate the exponential decrease of $J$ with
decreasing density stops at $n\sim d^{-1}$, because the potential screened
by the gates falls off rapidly at distances $x\gg d$.  Using the method
described in Appendix~\ref{sec:WKB}, one estimates
\begin{equation}
  \label{eq:J-Wigner-screened}
  J\sim E_F \left(\frac{nd^2}{a_B}\right)^{3/4}
        \exp\left(-\tilde\eta\sqrt\frac{d}{a_B}\right)
\end{equation}
at $a_B/d^2\ll n\ll d^{-1}$.  In the case of interaction potential
(\ref{eq:screened_interaction}) the constant $\tilde\eta
%=\sqrt{\pi/2}\,[\Gamma(1/8)/\Gamma(5/8)+\Gamma(5/8)/\Gamma(9/8)]
\approx8.49$.

The distance to the gate in quantum wire devices is typically large,
$d\gtrsim 10 a_B$, and most experiments are performed at densities well
above $a_B/d^{2}$.  However, if the density is reduced to $n\lesssim
a_B/d^{2}$, the Wigner crystal picture used in Sec.~\ref{sec:wigner} will
fail.  It is interesting to explore to what extent the conclusions of
Sec.~\ref{sec:wigner} will be affected.  To this end, let us now study the
limit $nd^2/a_B\to0$.

At $n\ll a_B/d^2$ the interaction between two particles at a typical
distance $n^{-1}$ is small compared to their kinetic energy $\sim E_F$.
On the other hand, when the distance between electrons is sufficiently
short, $|x|\lesssim n^{-1}(nd^2/a_B)^{1/3}\ll n^{-1}$, they experience
strong repulsion $V(x)\gtrsim E_F$.  Thus in the limit of low electron
density one can model the interaction potential
(\ref{eq:screened_interaction}) by short-range repulsion
\begin{equation}
  \label{eq:short_range}
  V(x)=\mathcal{V}\delta(x).
\end{equation}
The constant $\mathcal{V}$ should be chosen in such a way that the
scattering phase shift for two electrons interacting with potentials
$V(x)$ and $\mathcal{V}\delta(x)$ are identical.  For the model
(\ref{eq:screened_interaction}) this condition gives
\begin{equation}
  \label{eq:cal_V}
  \mathcal{V}\sim \frac{\hbar^2a_B}{md^2}
                  \exp\left(\tilde\eta\sqrt\frac{d}{a_B}\right).
\end{equation}
The exponentially large value of $\mathcal{V}$ reflects the fact that the
strong repulsion (\ref{eq:screened_interaction}) leads to almost perfect
backscattering of electrons off each other.  

At $\mathcal{V}\to\infty$ the electrons are separated by thin hard-core
potentials.  In this limit they can be viewed as distinguishable
particles, and the eigenvalues of energy become independent of the
electron spins.  The wavefunctions of the system essentially coincide with
the Slater determinants for spinless non-interacting fermions.  Upon
bosonization, the Hamiltonian $H_\rho$ of this system takes the form
(\ref{eq:H_rho}).  The plasmon velocity $s$ in this system is the Fermi
velocity of non-interacting electron gas of density $n$, which is twice
the Fermi velocity of non-polarized electron gas, $s=2v_F$.  Thus
according to Eq.~(\ref{eq:parameter_relations}) we have\footnote{One can
  easily check that the derivation of Eq.~(\ref{eq:parameter_relations})
  is valid not only for a Wigner crystal, but for any
  translation-invariant model of interacting electrons.} $K_\rho=1/2$.
Additional properties of this model were recently discussed in
Refs.~\onlinecite{cheianov,balents}.

At large finite $\mathcal{V}$ the electrons can change places as a result
of scattering, and the energy acquires a weak dependence on the spins.
This dependence can be deduced from the well-known properties of the
one-dimensional Hubbard model.  It has been shown by Ogata and
Shiba\cite{ogata,shiba} that at $U/t\to\infty$ the spin and charge
excitations of the Hubbard model are completely separated, with the
Hamiltonian of spin excitations taking the form of the Heisenberg model
(\ref{eq:spin-chain}).  The magnitude of the exchange constant in this
Hamiltonian was found\cite{shiba} to be
\begin{equation}
  \label{eq:J_Hubbard}
  J = \frac{4t^2}{U}n_e\left(1-\frac{\sin 2\pi n_e}{2\pi n_e}\right),
\end{equation}
where $n_e$ is the average number of electrons per site.  In the limit
$n_e\to0$ the Hubbard model is equivalent to an electron gas with
quadratic spectrum and point-like interaction (\ref{eq:short_range}).  The
limiting procedure can be performed by introducing infinitesimal lattice
period $a$ in the Hubbard model, identifying the parameters
$t=\hbar^2/2ma^2$, $U=\mathcal{V}/a$, $n_e=na$, and taking the limit
$a\to0$.  Applying this procedure to the formula (\ref{eq:J_Hubbard}), we
find
\begin{equation}
  \label{eq:J_point-like}
  J = \frac{2\pi^2}{3}\frac{\hbar^4n^3}{m^2\mathcal{V}}.
\end{equation}
Using the estimate (\ref{eq:cal_V}) of parameter $\mathcal{V}$ for the
interaction potential (\ref{eq:screened_interaction}), we find
\begin{equation}
  \label{eq:J_short-range}
  J\sim E_F \frac{nd^2}{a_B}
        \exp\left(-\tilde\eta\sqrt\frac{d}{a_B}\right).
\end{equation}
Note that the our results (\ref{eq:J-Wigner-screened}) and
(\ref{eq:J_short-range}) for the exchange constant are of the same order
of magnitude at $n=a_B/d^2$.

So far we have demonstrated that the description of the system in terms of
the Hamiltonian in spin-charge separated form $H=H_\rho+H_\sigma$, with
$H_\rho$ and $H_\sigma$ given by Eqs.~(\ref{eq:H_rho}) and
(\ref{eq:spin-chain}) is valid in two different regimes.  The first one is
the Wigner crystal state at electron densities in the range $a_B/d^2\ll n
\ll a_B^{-1}$, and the second is the low density limit $n\ll a_B/d^2$,
where the picture of point-like interactions (\ref{eq:short_range}) is
applicable.  One can show\cite{unpublished} that in fact this picture of
spin-charge separation holds at any density $n\ll a_B^{-1}$, including the
regime $n\sim a_B/d^2$.

The exchange constant $J$ in the effective spin chain Hamiltonian
(\ref{eq:spin-chain}) monotonically decreases as the electron density $n$
is lowered.  In the most interesting range of densities $d^{-1}\ll n\ll
a_B^{-1}$ the dependence of exchange on $n$ is exponential,
Eq.~(\ref{eq:J}).  At lower densities the dependence becomes a power-law
one.  Specifically, in the density ranges $a_B/d^2\ll n \ll d^{-1}$ and
$n\ll a_B/d^2$ one can use the estimates (\ref{eq:J-Wigner-screened}) and
(\ref{eq:J_short-range}), respectively.  

For the sake of simplicity, in the following sections we assume that the
electron density is in the range $a_B/d^2\ll n\ll a_B^{-1}$, and refer to
the electron system as a Wigner crystal.  However, all of our conclusions
remain valid at any densities $n\ll a_B^{-1}$, if the value of the
exchange constant $J$ is adjusted as discussed in this section.

\section{Conductance of a quantum wire with spin-charge separation}
\label{sec:separation}

The spin-charge separation has a profound effect on the conductance of
quantum wires.  Indeed, the electric field applied to the wire couples to
the electron charges and has no effect on spins.  As a result, the spin
degrees of freedom remain decoupled from charge ones, and the rather
complex form of the Hamiltonian $H_\sigma$ has no effect on the
conductance.  In this section we review the known results for the
conductance of a quantum wire with spin-charge separation.

\subsection{Infinite wire}
\label{sec:conductance_infinite_wire}

Conductance of an infinite Luttinger liquid is given by $G=2K_\rho e^2/h$.
This result was obtained\cite{kane,furusaki} by assuming that a weak
electric field is applied to a small part of the wire, and the conductance
was evaluated by using the Kubo formula.  In the following sections it
will be more convenient to evaluate the conductance of the Wigner crystal
in the regime of applied current.  It is therefore instructive to
reproduce the result $G=2K_\rho e^2/h$ in this approach.

Let us consider a quantum wire whose charge dynamics are described by the
Hamiltonian (\ref{eq:H_rho}), and enforce the current $I=I_0\cos\omega t$
at $x=0$.  By doing so we impose a boundary condition upon the charge
field $\phi_\rho(0,t)$.  Indeed, the bosonization expression for the
electric current is $I=e\frac{\sqrt2}{\pi}\dot\phi_\rho$.  (In the case of
a Wigner crystal, this can be checked by using
Eq.~(\ref{eq:field_identities}) and the definition $I=en\dot u$ of current
in terms of the velocity $\dot u$ of the crystal.)  Thus the field
$\phi_\rho$ satisfies the condition
\begin{equation}
  \label{eq:boundary_condition_phi_rho}
  \phi_\rho(0,t) = \frac{\pi}{\sqrt2}\, q(t),
\end{equation}
where the function
\begin{equation}
  \label{eq:q}
  q(t) = \frac{I_0}{e\omega}\sin \omega t.
\end{equation}
is related to the current as $I=e\dot q$ and has the meaning of the number
of electrons that passed through point $x=0$ at time $t$.

By imposing a time-dependent boundary condition
(\ref{eq:boundary_condition_phi_rho}) we drive the system with an external
oscillating force.  This leads to emission of plasmon waves and
dissipation of the energy from the driving force to the infinite Luttinger
liquid.  We will find the resistance of the wire $R_\rho$ by evaluating
the energy $W$ dissipated in unit time and comparing the result with the
Joule heat law $W=\frac12 I_0^2 R_\rho$.  We present a formal derivation
in Appendix~\ref{sec:quantum_resistance}; here we limit ourselves to a
simple semiclassical argument.

Solving the Hamilton equations with Hamiltonian (\ref{eq:H_rho}) and
boundary condition (\ref{eq:boundary_condition_phi_rho}) we
find\footnote{The asymptotics at $x\to\pm\infty$ in
  Eq.~(\ref{eq:plasmon_wave}) are chosen to correspond to outgoing
  plasmon wave.}
\begin{subequations}
 \label{eq:plasmon_wave}
\begin{eqnarray}
  \label{eq:classical_solution}
  \phi_\rho(x,t) &=& \frac{\pi I_0}{\sqrt2 e\omega}
                     \sin\omega(t-|x|/u_\rho),
  \\
  \Pi_\rho(x,t)  &=& \frac{I_0}{\sqrt2 e K_\rho u_\rho}
                     \cos\omega(t-|x|/u_\rho).
\end{eqnarray}
\end{subequations}
Substituting this solution back into Eq.~(\ref{eq:H_rho}), we find the
following expression for the time-averaged energy density in the Luttinger
liquid,
\begin{equation}
  \label{eq:energy_density}
  \langle E\rangle_t = \frac{\pi\hbar}{4e^2}\,  \frac{I_0^2}{K_\rho u_\rho}.
\end{equation}
The plasmon wave (\ref{eq:plasmon_wave}) carries the energy $\langle
E\rangle_t$ at speed $u_\rho$ in two directions.  Thus the total energy
dissipated into plasmon waves in unit time is given by $W=2 u_\rho \langle
E\rangle_t$.  Comparing this result with $W=\frac12 I_0^2 R_\rho$, we find
the resistance
\begin{equation}
  \label{eq:R_rho_renormalized}
  R_\rho = \frac{h}{2K_\rho e^2},
\end{equation}
in agreement with the result for the conductance found in
Ref.~\onlinecite{kane,furusaki}.

\subsection{Finite-length quantum wire between two non-interacting leads}
\label{sec:maslov}

The result $G=2K_\rho e^2/h$ indicates that in a quantum wire with
repulsive interactions conductance should be below the quantized value
$2e^2/h$.  Furthermore, it is expected to decrease as the electron density
$n$ is lowered.  However, the experiments consistently show perfect
quantization\footnote{Experiments with cleaved edge overgrowth
  wires,\cite{yacoby} show quantization below $2e^2/h$, but this value is
  still independent of the electron density.  The reduction of the
  quantized value was attributed to the imperfect coupling of the wire to
  the leads.\cite{picciotto}} of conductance at $2e^2/h$ in a broad range
of $n$.

This controversy was resolved\cite{maslov,ponomarenko,safi} by noticing
that instead of an infinite quantum wire, the experiments study transport
through a finite-length wire connecting two bulk leads.  Since the leads
are not one-dimensional, their properties are not adequately described by
the Luttinger liquid model (\ref{eq:separation})--(\ref{eq:H_sigma}).
Instead, the electrons in the leads are expected to be in a Fermi liquid
state.  

To find the conductance of such devices, one can
model\cite{maslov,ponomarenko,safi} the leads connected to the wire by two
semi-infinite non-interacting wires.  In this model the system remains
one-dimensional, but the interactions are non-vanishing only in the
central part of the system.  The length $L$ of the interacting part is
identified with the length of the wire.  Assuming that the interactions
fall off gradually at $x\to\pm\infty$, one can neglect the back-scattering
of electrons from the interacting region.  In this limit the charge
dynamics are still described by the Hamiltonian (\ref{eq:H_rho}), but the
parameters $K_\rho$ and $u_\rho$ become functions of coordinate $x$.

The measurements of dc conductance in experiments are conducted at very
low frequencies $\omega\ll u_\rho/L$.  Thus the wavelength of the plasmons
emitted in the system is much greater than the length of the interacting
region $L$.  Consequently, one should use the value of the parameter
$K_\rho$ in Eq.~(\ref{eq:R_rho_renormalized}) taken at $x\to\pm\infty$,
i.e., in the non-interacting region, where $K_\rho=1$.  Thus the
resistance (\ref{eq:R_rho_renormalized}) of the device becomes
\begin{equation}
  \label{eq:R_rho}
  R_\rho = \frac{h}{2 e^2},
\end{equation}
restoring the perfect quantization.  Careful
treat\-ments\cite{maslov,ponomarenko,safi} of the problem lead to the same
conclusion.

\section{Violation of spin-charge separation in quantum wire devices}
\label{sec:nonseparation}

As we saw in Sec.~\ref{sec:separation}, the inhomogeneity of the system
caused by coupling of the wire to the leads changes the conductance from
$2K_\rho e^2/h$ to $2e^2/h$.  This conclusion was derived from
consideration of the charge excitations only, as the spin degrees of
freedom were assumed to be completely separated.  We now turn to the
effect of the inhomogeneity on the spin part $H_\sigma$ of the
Hamiltonian.

We assume that the central part of the wire contains a purely
one-dimensional electron system at low density $n\ll a_B^{-1}$, so that
the Wigner crystal model is appropriate.  The wire is also assumed to be
smoothly connected to the leads, where the effective interactions are
weak.  This is due to several effects.  First, the electron density grows
as one moves away from the wire into the leads.  This effectively reduces
the interaction strength, as the latter is characterized by parameter
$(na_B)^{-1}$.  In addition, the wire becomes wider when it couples to the
leads.  As a result, when two electrons arrive at the same coordinate $x$
along the wire, they are no longer as close to each other as in the middle
of the wire.  This reduces the strength of interactions between 1D
electrons.  The two mechanisms have very similar effect on the Hamiltonian
$H_\sigma$.  For simplicity, in the following we limit our discussion to
the effect of inhomogeneous electron density.

Following the ideas of Refs.~\onlinecite{maslov,ponomarenko,safi}, we
model the wire connected to the leads by an inhomogeneous 1D system.  The
main source of inhomogeneity is the dependence $n(x)$ of the electron
density on position.  We assume that the density takes a constant value
$n(x)=n$ inside the wire, i.e., at $|x|<L/2$, and gradually grows to a
very large value $n_\infty\gg a_B^{-1}$ at $x\to\pm\infty$.

In experimental devices the dependence of electron density on the
coordinate along the wire is caused by inhomogeneity of the external
confining potential.  Apart from changing the electron density, the
external potential may also lead to backscattering of electrons in the
wire.  In a sufficiently long wire such processes may greatly suppress the
conductance at low temperature.\cite{kane,furusaki} In the Wigner crystal
picture this phenomenon is interpreted as pinning of the crystal by the
external potential.\cite{ruzin} On the other hand, the best available
experiments show good quantization of conductance, indicating that the
backscattering remains negligible.  This is most likely the result of
smoothness of the confining potential.  Indeed, the backscattering
involves the change of the electron wavevector by $2k_F$.  Thus an
external potential that is smooth at the scale of interparticle distance
$n^{-1}$ will cause exponentially weak backscattering.  In this paper we
assume that the external potential is sufficiently smooth, so that the
backscattering can be neglected.

Under the above conditions the low-energy properties of the system may be
described by the bosonized Hamiltonian
(\ref{eq:separation})--(\ref{eq:H_sigma}), but with position-dependent
parameters $u_{\rho,\sigma}$, $K_{\rho,\sigma}$, $g_{1\perp}$.  In this
paper we assume that the temperature is small compared with the bandwidth
$D_\rho\sim \hbar n u_\rho$ of the Hamiltonian $H_\rho$, so that the
discussion of the effect of the charge modes on the conductance presented
in Sec.~\ref{sec:separation} is valid.  On the other hand, we will be
interested in the case of temperature comparable with the bandwidth
$D_\sigma\sim J$ of the Hamiltonian $H_\sigma$.  In this regime the
bosonized version (\ref{eq:H_sigma}) of $H_\sigma$ is not applicable, and
one should instead use the Heisenberg model (\ref{eq:spin-chain}).

Since the exchange constant (\ref{eq:J}) strongly depends on the electron
density $n(x)$, the parameter $J$ in Eq.~(\ref{eq:spin-chain}) should also
be considered position-dependent.  In particular, the strength of the
exchange coupling between the two spins at the neighboring sites $l$ and
$l+1$ of the Wigner lattice is a function of the coordinate $x_l$ of the
$l$-th electron: $J=J(x_l)$.  It is important to note that in the presence
of electric current $I$ the Wigner lattice moves, so the coordinate $x_l$
of the $l$-th lattice site depends not only on $l$, but also on time.

The time dependence of $x_l$ can be accounted for by noting that if during
the time interval $t$ a number $q(t)$ of electrons have moved from the
left lead to the right one, the $l$-th site of the lattice has shifted to
the $(l+q)$-th position.  Thus the time dependence of the positions of the
lattice sites can be accounted for by replacing $l\to l+q(t)$, and the
Hamiltonian $H_\sigma$ takes the form
\begin{equation}
  \label{eq:spin-chain_inhomogeneous}
  H_\sigma = \sum_l J[l+q(t)]\,
             {\bm S}_{l}\cdot{\bm S}_{l+1}.
\end{equation}
Note that in this approximation the electric current $I=e\dot q(t)$ is
assumed to be uniform throughout the wire.  This is true in the dc limit
$\omega \ll u_\rho/L$.

It is important to note that the form (\ref{eq:spin-chain_inhomogeneous})
of the Hamiltonian $H_\sigma$ violates the spin-charge separation.
Indeed, the coupling between the spins depends on the amount of charge
that passed through the wire, which is related to the field $\phi_\rho$,
see Eq.~(\ref{eq:boundary_condition_phi_rho}).  As a result, the
conductance of a quantum wire connected to bulk leads may be affected by
the spin excitations.\footnote{It is worth mentioning that the violation
  of the spin-charge separation in Eq.~(\ref{eq:spin-chain_inhomogeneous})
  is due to the inhomogeneity of the system.  In the homogeneous case
  $J=\rm const$, and the spin-charge separation is restored.}

To find the effect of spin subsystem on the conductance, one could
substitute the expression (\ref{eq:boundary_condition_phi_rho}) for $q(t)$
into Eq.~(\ref{eq:spin-chain_inhomogeneous}), and consider the complete
Hamiltonian $H_\rho+H_\sigma$ without relying on spin-charge separation.
In this approach one needs to add to the Hamiltonian a term describing the
applied bias, and evaluate the electric current.  However, it is more
convenient to treat the current $I(t)$ in the wire as an external
parameter.  In this case $q(t)$ is also a parameter, and the Hamiltonians
$H_\rho$ and $H_\sigma$ still commute.  The only consequence of the
violation of spin-charge separation in this approach is the dependence of
$H_\sigma$ on the current $I(t)$.

The presence of an oscillating parameter (\ref{eq:q}) in the Hamiltonian
(\ref{eq:spin-chain_inhomogeneous}) may lead to creation of spin
excitations.  Using the approach of Sec.~\ref{sec:separation}, we will
calculate the energy dissipated into spin excitations in unit time.  In
the limit of weak current, the dissipation is found in the second order of
the perturbation theory in the amplitude $I_0$ of current oscillations.
Thus in addition to the plasmon result for the energy $W$ dissipated in
unit time, we will obtain a similar contribution of the spin modes:
\begin{equation}
  \label{eq:dissipation}
  W=\frac12 I_0^2 R_\rho + \frac12 I_0^2 R_\sigma.
\end{equation}
Comparing this result with the Joule heat law $W=\frac12 I_0^2 R$, we
conclude that the resistance $R$ of the wire is given by sum of two
independent contributions,
\begin{equation}
  \label{eq:Kirchoff}
  R = R_\rho + R_\sigma.
\end{equation}
Note that the first term in this expression is already known,
Eq.~(\ref{eq:R_rho}).  The second term is discussed in
Sec.~\ref{sec:conductance}.

It is interesting to point out that the result (\ref{eq:Kirchoff}) may be
interpreted as a total resistance of the charge and spin subsystems
connected in series, whereas naively one might expect a parallel
connection.  The reason is that the spins do not directly respond to the
applied voltage, as required for the latter interpretation.  Instead, the
spin subsystem responds to the electric current.  Thus the Hamiltonians
$H_\rho$ and $H_\sigma$ become independent in the regime of applied
current, in analogy with the problem of two independent resistors
connected in series.  A result similar to Eq~(\ref{eq:Kirchoff}) has been
obtained for the resistivity of two-dimensional strongly interacting
systems in Ref.~\onlinecite{ioffe}.

\section{Spin contribution to the resistance}
\label{sec:conductance}

To find the contribution $R_\sigma$ of the spin subsystem to the
resistance of the device, we study the dissipation of energy into spin
excitations caused by the time dependence of the Hamiltonian
(\ref{eq:spin-chain_inhomogeneous}).  We start by performing the
Jordan-Wigner transformation (\ref{eq:Jordan-Wigner}) and converting the
Hamiltonian to the fermionic form
\begin{eqnarray}
    H_\sigma &=& \frac12 \sum_l J[l+q(t)]\biggl[
               \left(a_l^\dagger a_{l+1} + a_{l+1}^\dagger a_{l}\right)
\nonumber\\
             && + 2\left(a_l^\dagger a_l + \frac12\right)
                       \left(a_{l+1}^\dagger a_{l+1} + \frac12\right)
                 \biggl].
  \label{eq:fermionic_time-dependent}
\end{eqnarray}
In the absence of the external magnetic field the average $z$-component of
the spin at every site of the lattice must vanish.  Thus according to
Eq.~(\ref{eq:Jordan-Wigner}) the occupation of each site is $\langle
a^\dagger_l a_l\rangle =\frac12$.  This means that the Fermi level is in
the middle of the band, $\mu=0$.

The exchange $J[y]$ strongly depends on the position $y$.  Inside the wire
the electron density is low, $na_B\ll1$, and the exchange is exponentially
small, Eq.~(\ref{eq:J}).  As the wire connects to the bulk leads, the
density $n(x)$ begins to grow.  At $na_B\sim 1$ the exchange $J$ becomes
of the order of the Fermi energy, see Eqs.~(\ref{eq:J}),
(\ref{eq:prefactor}).

Strictly speaking, the Wigner crystal picture is valid only at $na_B\ll
1$, that is as long as $J\ll E_F$.  On the other hand, we will be
interested in the properties of the system at low energies $D\ll E_F$.
Thus at $J\sim E_F$ when the Wigner crystal picture fails, we are only
concerned with the energy scales much lower than $J$.  As we saw in
Sec.~\ref{sec:spin_charge}, at those scales one can use the bosonized
Hamiltonian $(\ref{eq:H_sigma})$ regardless of the applicability of the
Wigner crystal description.  Thus we can ignore the difference between the
Wigner crystal and weakly interacting electron gas at large density $n\gg
a_B$, and simply assume that in the leads the exchange $J$ saturates at
$J_\infty\sim E_F$.

\begin{figure}
 \resizebox{.42\textwidth}{!}{\includegraphics{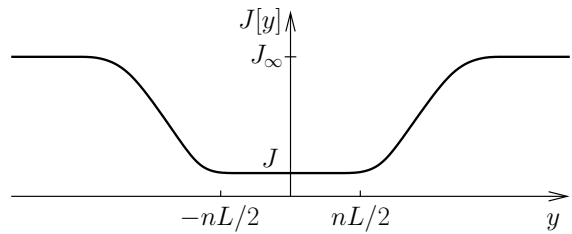}}
\caption{\label{fig:plot}Scketch of the dependence $J[y]$ in our model.  
  Inside the wire, $|y|<nL/2$, the exchange $J$ is exponentially small,
  Eq.~(\ref{eq:J}).  As one moves toward the leads, $J$ grows, and at
  $y\to\infty$ it saturates at $J_\infty\sim E_F$.}
\end{figure}

The properties of the function $J[y]$ can thus be summarized as follows,
\begin{equation}
  \label{eq:J[y]}
  J[y]=\left\{
      \begin{array}{ll}
         J \ll E_F,          & \mbox{at $|y|<nL/2$,}\\[1ex]
         J_\infty\sim E_F,   & \mbox{at $|y|\to\pm\infty$,}
      \end{array}
       \right.
\end{equation}
see Fig.~\ref{fig:plot}.  Note that $y$ is the coordinate on the Wigner
lattice.  Since we consider the limit of very smooth confining potential,
all the physical quantities change very little at the interparticle
distance.  We therefore assume that $J[y]$ is a slowly varying function:
$|dJ/dy|\ll J[y]$.

\subsection{XY model}
\label{sec:XY}

The Hamiltonian (\ref{eq:fermionic_time-dependent}) describes a system of
strongly interacting fermions.  As a first approximation we will simplify
the problem by neglecting the interactions between the fermions,
\begin{equation}
    H_\sigma^{XY} = \frac12 \sum_l J[l+q(t)]
               \left(a_l^\dagger a_{l+1} + a_{l+1}^\dagger a_{l}\right).
  \label{eq:XY}
\end{equation}
This Hamiltonian corresponds to the fermionized version of the XY model of
a spin chain, in which the coupling of the $z$-components of spin
operators is neglected.  This approximation violates the SU(2) symmetry of
the problem, and is therefore rather crude.  On the other hand, the
resistance $R_\sigma$ can be found exactly for model (\ref{eq:XY}), and
the result will provide considerable insight into the properties of model
(\ref{eq:fermionic_time-dependent}).

Hamiltonian (\ref{eq:XY}) represents an inhomogeneous version of the
the tight-binding model of lattice fermions.  In the uniform case,
$J[y]=\rm const$, the spectrum is well known,
\begin{equation}
  \label{eq:spectrum_XY}
  \epsilon(k) = J \sin k,
\end{equation}
where the wavevector $k$ is measured from $k_F=\pi/2$.  One can either
assume that $k$ varies in the interval $-\pi<k<\pi$, or choose $0<k<\pi$
and treat Eq.~(\ref{eq:spectrum_XY}) as spectra of two branches of
excitations, the particles and holes.

In the absence of electric current in the wire one can omit $q(t)$ in the
Hamiltonian (\ref{eq:XY}) and view it as a tight-binding model with slowly
varying bandwidth $2J[y]$.  In the leads the bandwidth $2J_\infty$ is very
large; it narrows down to a very small value $2J$ in the wire,
Eq.~(\ref{eq:J[y]}).  The particles moving toward the wire in one of the
leads cross to the other lead if their energies are below the small
exchange $J$ in the wire; the particles with $\epsilon>J$ are reflected.

In the presence of the electric current $I$, the constriction of the band
in the Hamiltonian (\ref{eq:XY}) moves with respect to the lattice with
velocity $\dot q = I/e$.  The particles reflected from the moving
constriction change their energy.  These processes lead to the dissipation
of energy and contribute to the resistance $R_\sigma$.

For non-interacting fermions, the problem of evaluating the energy $W$
dissipated in unit time by a moving scatterer can be solved for arbitrary
reflection coefficient $\mathcal{R}(\epsilon)$, see
Appendix~\ref{sec:dissipation_XY}.  Here we find $W$ in the semiclassical
limit, which is valid for very slowly varying bandwidth $J[y]$, when
$\mathcal{R}(\epsilon)= \theta(\epsilon-J)$.

In the limit of slowly varying $J[y]$ one can apply the result
(\ref{eq:spectrum_XY}) for the spectrum of particles at every point in
space, and treat the excitations as classical particles with energy
\begin{equation}
  \label{eq:energy_classical}
  H(y,p,t) = J[y+vt] \sin \frac{p}{\hbar}.
\end{equation}
Here $y$ is the coordinate of the particle, $p$ is its momentum, and
$v=I/e$ is the velocity of the constriction.  For simplicity we will
consider the case of dc current, $I=\rm const$.   To find the linear
conductance of the quantum wire, one can limit oneself to the case of very
small current, and assume $v\ll T/\hbar, J/\hbar$.

The time-dependent energy (\ref{eq:energy_classical}) should be treated as
a Hamilton function, and the trajectory of the particle can, in principle,
be found by solving the classical Hamilton equations.  One can easily
check that the quantity
\begin{equation}
  \label{eq:integral-of-motion}
  E(y,p,t) = H(y,p,t) + pv
\end{equation}
is an integral of motion.  It has the meaning of energy of the particle in
the frame moving at the speed of the constriction.

A particle with energy $\epsilon\sim T$ moving in the right direction has
a very low momentum $p$ when it is in the leads, $p/\hbar
=\epsilon/J_\infty \sim T/E_F \ll 1$.  Thus its integral of motion
$E(y,p,t)=\epsilon$.  As the particle approaches the constriction, its
momentum increases, so that $E$ retains its value despite the decrease of
the bandwidth $J$.  At small $v$ the maximum allowed value of $E$ in the
wire is reached at $p=\pi\hbar/2$ and equals $E_\textrm{max} = J
+\frac{\pi}{2} \hbar v$.  Thus at $\epsilon<J +\frac{\pi}{2} \hbar v$ the
right-moving particle moves from the left lead to the right one, and its
energy $\epsilon$ remains unchanged.  If the energy $\epsilon$ exceeds $J
+\frac{\pi}{2} \hbar v$, the particle cannot enter the wire.  When its
momentum reaches $\pi\hbar/2$ at a point to the left of the constriction,
the particle is reflected.  Deep in the left lead its momentum is very
close to $\pi\hbar$.  Due to the $pv$ term in integral of motion
(\ref{eq:integral-of-motion}), its energy $H$ decreases to $\epsilon -
\pi\hbar v$.

Similarly, since a left-moving particle in the right lead with energy
$\epsilon$ has momentum very close to $\pi\hbar$, its integral of motion
(\ref{eq:integral-of-motion}) is $E=\epsilon+\pi\hbar v$.  The condition
$E<E_\textrm{max}$ for transmission through the constriction for such
particles means $\epsilon<J -\frac{\pi}{2} \hbar v$.  As the particle
reaches the left lead, the momentum is again near $\pi\hbar$, i.e.,
conservation of $E$ results in conservation of energy $H=\epsilon$.   On
the other hand, particles with energies $\epsilon>J -\frac{\pi}{2} \hbar
v$ are reflected back to the right lead, and their momentum on the
right-moving branch is near $p=0$.  Thus the energy of these particles
increases from $\epsilon$ to $\epsilon+\pi\hbar v$.

%% A similar argument shows that left-moving particles with energies
%% $\epsilon< J -\frac{\pi}{2}\hbar v$ in the right lead cross the
%% constriction and arrive at the left lead without change of energy.
%% Particles with $\epsilon> J - \frac{\pi}{2}\hbar v$ return to the right
%% lead with increased energy $\epsilon + \pi\hbar v$.  \textbf{Expand?}

To summarize, the particles in the leads with energies $\epsilon<J -
\frac{\pi}{2}\hbar v$ cross the constriction region without change of
energy.  The particles with energies $\epsilon>J + \frac{\pi}{2}\hbar v$
are always reflected by the constriction.  The ones in the left lead
decrease their energy by $\pi\hbar v$, while the ones in the right lead
increase their energy by the same amount, so that these contributions to
the total energy of the system compensate each other.  Finally, in the
narrow range of energies $J - \frac{\pi}{2}\hbar v <\epsilon< J +
\frac{\pi}{2}\hbar v$ the right-movers go through the constriction without
change of energy, whereas the left-movers reflect back to the right lead
with energy gain $\pi\hbar v$.

The total current of left moving particles and holes in the narrow energy
interval of width $\pi\hbar v$ near $\epsilon=J$ is given by $\delta\dot
N=(2/h) (\pi\hbar v) f(J)$, where
$f(\epsilon)=1/(e^{\epsilon/T}+1)$ is the Fermi function.  Thus the
total energy transferred to the spin excitations in unit time is
\begin{equation}
  \label{eq:W_XY}
  W^{XY} = \pi\hbar v^2 f(J)=I^2\frac{\pi\hbar}{e^2} f(J).
\end{equation}

Comparing this result with the Joule heat law $W=I^2R$, we obtain the spin
contribution to the resistance
\begin{equation}
  \label{eq:R_XY}
  R_\sigma^{XY} = \frac{h}{2e^2} f(J).
\end{equation}
At low temperature $T\ll J$ most of the particles have energies below $J$
and pass through the constriction elastically.  Only an exponentially
small fraction of particles are reflected and contribute to the
dissipation.  Thus the result (\ref{eq:R_XY}) is exponentially small at
low temperatures, $R_\sigma^{XY}\simeq (h/2e^2) e^{-J/T}$.  As the
temperature is increased, a greater fraction of the particles are
reflected by the constriction, and the $R_\sigma^{XY}$ increases.  In the
limit $T/J\to\infty$ all the particles are reflected, and the resistance
saturates at $R_\sigma^{XY}=h/4e^2$.

In this section we studied the simplified XY model, in which the
$z$-component of coupling in the Hamiltonian
(\ref{eq:spin-chain_inhomogeneous}) was neglected.  Thus the result
(\ref{eq:R_XY}) cannot be applied directly to the problem of conductance
of a quantum wire in the Wigner crystal regime.  However, much of the
physics leading to Eq.~(\ref{eq:R_XY}) can be carried over to the case of
the isotropic model (\ref{eq:spin-chain_inhomogeneous}).

\subsection{Isotropic coupling}
\label{sec:isotropic}

The problem of the isotropic inhomogeneous Heisenberg spin chain
(\ref{eq:spin-chain_inhomogeneous}) is far more complicated than that of
the XY model (\ref{eq:XY}).  However, it can still be somewhat simplified
by assuming that $J[y]$ is a very slowly varying function.  Then each
moderately long section of the spin chain can be approximated by the
homogeneous Heisenberg model.  The latter allows for exact
solution\cite{bethe} by means of Bethe ansatz.  The low-energy excitations
of the isotropic Heisenberg spin chain are spinons with energy
spectrum\cite{des_cloizeaux,faddeev}
\begin{equation}
  \label{eq:spectrum_spinons}
  \epsilon(k) = \frac{\pi J}{2} \sin k.
\end{equation}
Although the spinons do not obey Fermi statistics, the similarity between
Eq.~(\ref{eq:spectrum_spinons}) and the spectrum (\ref{eq:spectrum_XY}) of
the excitations of XY model enables us to find the temperature dependence
of $R_\sigma$ at $T\ll J$.

Indeed, most of the discussion leading to Eq.~(\ref{eq:R_XY}) did not rely
on Fermi statistics of the excitations.  One can apply the arguments of
Sec.~\ref{sec:XY} to the problem of scattering of spinons by the constriction
of the band in the wire.  In particular, one concludes that spinons with
energies below $\pi J/2$ pass through the constriction without scattering
and do not change their energy.  Thus the dissipation is exponentially
small at $T\ll J$, and one finds $R_\sigma\propto\exp(-\pi J/2T)$.

Since the occupation of states with high energy $\epsilon=\pi J/2$ at low
temperature is exponentially small and independent of statistics, on can
naively expect the resistance $R_\sigma$ to be given by the
low-temperature asymptotics of Eq.~(\ref{eq:R_XY}) upon replacement $J\to
\pi J/2$.  Then one obtains
\begin{equation}
  \label{eq:R_sigma-low-T}
  R_\sigma = R_0 \exp\left(-\frac{\pi J}{2T}\right).
\end{equation}
This approach gives the prefactor $R_0=h/2e^2$.

Unfortunately the analogy between spinons and fermion excitations of the
XY model does not enable one to find the prefactor in
Eq.~(\ref{eq:R_sigma-low-T}).  Unlike the excitations of the XY model, the
spinons interact with each other, and the energy of a spinon is affected
by the presence of other spinons.  In the limit $T\to 0$ the density of
other spinons is small, and the energy is given by
Eq.~(\ref{eq:spectrum_spinons}).  At finite $T$ the result
(\ref{eq:spectrum_spinons}) may acquire a small correction.  The exponent
of Eq.~(\ref{eq:R_sigma-low-T}) is determined by the maximum energy of a
spinon $\pi J/2$.  Even a small correction to this energy may affect the
prefactor $R_0$.

At high temperature $T\gg J$ the resistance contribution $R_\sigma$
evaluated within the XY model approximation saturates, because in this
regime all the excitations are reflected by the constriction.  This
feature is preserved in the model with isotropic coupling, as at $J\to 0$
the spin excitations cannot propagate through the wire.  The saturation
value of $R_\sigma$ at high temperatures can be found by noticing that in
the part of the system away from the constriction, where $J[l]\simeq
J_\infty \gg T$, one can still bosonize the Hamiltonian
(\ref{eq:fermionic_time-dependent}) and use the form (\ref{eq:H_sigma}).
Inside the constriction the bosonization is not applicable, and this
region is modeled by imposing a boundary condition on the bosonic field
$\phi_\sigma$ corresponding to the fact that there is no spin propagation
between the regions of large positive and negative~$l$.

The form of the boundary condition can be deduced by replacing the
constriction (\ref{eq:J[y]}) of the bandwidth $D_\sigma \sim J[l]$ in the
Hamiltonian (\ref{eq:fermionic_time-dependent}) with a high potential
barrier for the fermions.  The barrier is modeled by a large
backscattering term $\upsilon(a_L^\dagger a_R + a_R^\dagger a_L)$ at site
$l=-q(t)$.  Upon the bosonization transformation
(\ref{eq:bosonization_spin-chain}) this term becomes
$-\tilde\upsilon\cos[\sqrt2\phi_\sigma(y)-2k_Fy]|_{y=-q(t)}$, with $k_F$
on the lattice being $\pi/2$.  Since this scattering term is very strong,
$\tilde\upsilon\to\infty$, it pins the field $\phi_\sigma(-q(t),t)$ to the
value $-\sqrt2 k_F q(t) = -(\pi/\sqrt2) q(t)$.  This time-dependent
boundary condition leads to the emission of spin waves, in analogy with
Sec.~\ref{sec:conductance_infinite_wire}, where the boundary condition
(\ref{eq:boundary_condition_phi_rho}) gave rise to plasmon waves
(\ref{eq:plasmon_wave}).  In the limit of weak current, $I\sim e\omega q
\to 0$, the wavelength of the spin waves $\sim J_\infty/\hbar\omega$ is
much larger than $q(t)$, and instead of imposing the boundary condition at
$y=-q(t)$ one can impose it at $y=0$.  Then the boundary condition becomes
\begin{equation}
  \label{eq:boundary_condition_phi_sigma}
  \phi_\sigma(0,t) = -\frac{\pi}{\sqrt2}\, q(t).
\end{equation}

Note that up to inessential negative sign
Eq.~(\ref{eq:boundary_condition_phi_sigma}) is identical to the boundary
condition (\ref{eq:boundary_condition_phi_rho}).  The respective
Hamiltonians (\ref{eq:H_rho}) and (\ref{eq:H_sigma}) are also essentially
identical at low energies, as the sine-Gordon term is irrelevant.  One can
therefore carry over the results of
Sec.~\ref{sec:conductance_infinite_wire} for the dissipation of energy
into plasmon waves and the resulting contribution to the resistance.
Adapting Eq.~(\ref{eq:R_rho_renormalized}) to the parameters of
Hamiltonian (\ref{eq:H_sigma}), we find
\begin{equation}
  \label{eq:R_sigma_renormalized}
  R_\sigma = \frac{h}{2K_\sigma e^2}.
\end{equation}
In the dc limit the frequency of the driving force $\omega\to0$, and the
wavelength of the spin waves is very long.  Thus the parameter $K_\sigma$
in Eq.~(\ref{eq:R_sigma_renormalized}) is taken at large distances from
the constriction, where the SU(2) symmetry demands $K_\sigma=1$.
Consequently the spin contribution to the resistance in the model with
isotropic coupling is given by
\begin{equation}
  \label{eq:R_sigma}
  R_\sigma = \frac{h}{2 e^2}.
\end{equation}
On the other hand, the XY model (\ref{eq:XY}) does not possess the SU(2)
symmetry, and the bosonization procedure
(\ref{eq:bosonization_spin-chain}) gives the quadratic part of Hamiltonian
(\ref{eq:H_sigma}) with $K_\sigma=2$.  Then
Eq.~(\ref{eq:R_sigma_renormalized}) predicts $R_\sigma^{XY}=h/4e^2$, in
agreement with $T\gg J$ asymptotics of Eq.~(\ref{eq:R_XY}).

\section{Discussion of the results}
\label{sec:discussion}

The quantity most commonly measured in experiments with quantum wire
devices is the linear conductance.  In our theory its value is given by
\begin{equation}
  \label{eq:conductance_result}
  G = \frac{1}{R_\rho+R_\sigma},
\end{equation}
c.f. Eq.~(\ref{eq:Kirchoff}).  The contributions $R_\rho$ and $R_\sigma$
to the resistance of the wire are determined by the properties of the
charge and spin excitations of the system, respectively.  

Throughout this paper we consider the case of relatively low temperature,
$T\ll D_\rho\sim\hbar n u_\rho$.  In this regime the contribution of the
charge modes is well known: $R_\rho=h/2e^2$ (see also
Sec.~\ref{sec:maslov}).  Raising the temperature above $D_\rho$ leads to
thermal smearing of conductance plateaus.  No interesting electron
correlation effects are expected in this case.

At not too low electron density $n\gtrsim a_B^{-1}$ the bandwidth
$D_\rho\sim D_\sigma\sim E_F$ is the only relevant energy scale of the
problem.  Then at $T\ll E_F$ the contribution $R_\sigma$ vanishes, and the
conductance takes the well-known quantized value $G=2e^2/h$.  On the other
hand, in the interesting case of low density $n\ll a_B^{-1}$ another
energy scale, the exchange constant $J$, appears in the problem.  This
scale is exponentially small, Eq.~(\ref{eq:J}); in particular, $J\ll
D_\rho$.  In the limit of low temperature $T\to0$ the contribution
$R_\sigma$ still vanishes.  More specifically, at $T\ll J$ we predict
activated temperature dependence (\ref{eq:R_sigma-low-T}) of $R_\sigma$,
with activation temperature $\pi J/2$.  At higher temperatures $R_\sigma$
grows, and at $T\gg J$ it saturates at the universal value
$R_\sigma=h/2e^2$, see Eq.~(\ref{eq:R_sigma}).  Combining these results
with Eq.~(\ref{eq:conductance_result}), we obtain
\begin{equation}
  \label{eq:G-high-T}
  G=\frac{e^2}{h},
    \quad
    J\ll T \ll D_\rho.
\end{equation}
This is our main result.  It corresponds to an additional quantized
plateau of conductance of a quantum wire at low electron density.  The
value of the conductance at this plateau is exactly one half of the
quantized conductance $2e^2/h$.

The plateaus of conductance at $e^2/h$ have been observed at low electron
densities in several experiments.\cite{bkane,thomas2,reilly1,reilly2} The
authors of Refs.~\onlinecite{bkane,thomas2,reilly1,reilly2} attributed
this feature to the spontaneous spin polarization in quantum wires.  This
interpretation contradicts to the theorem by Lieb and Mattis,\cite{mattis}
stating that the ground state of a 1D electron system cannot be
spin-polarized in the absence of magnetic field.  One can hypothesize that
the ferromagnetism in quantum wires is possible because the electrons are
not truly one-dimensional; however to the best of our knowledge, no such
theory is available at this time.  In our theory the spin structure of the
Wigner crystal state is described by the Heisenberg model
(\ref{eq:spin-chain}) with positive exchange constant $J$, corresponding
to antiferromagnetic coupling.  Thus the ground state of the Wigner
crystal in not spin-polarized, in agreement with the theorem.\cite{mattis}

The temperature dependence of the conductance of a quantum wire device
obtained in this paper is similar to the behavior observed in experiments
on 0.7 structure in quantum point
contacts.\cite{thomas,thomas1,cronenwett,kristensen1,kristensen} In
agreement with experiments, conductance (\ref{eq:conductance_result})
remains $2e^2/h$ at $T\to 0$, but develops a negative correction at finite
temperature.  The activated temperature dependence of the correction
following from Eq.~(\ref{eq:R_sigma-low-T}) is consistent with the
measurements of Ref.~\onlinecite{kristensen}.  At high temperature the
correction saturates, and the conductance develops a new plateau.
Contrary to the experiments,\cite{thomas,thomas1,cronenwett} this plateau
is at one half of the quantized value $2e^2/h$, rather than at $0.7\times
(2e^2/h)$.  The relation between the plateau at $e^2/h$ and the 0.7
structure was studied experimentally in Ref.~\onlinecite{reilly1}.  It was
found that the quasi-plateau at $0.7\times (2e^2/h)$ is observed in short
wires, whereas in longer wires it shifts toward $e^2/h$.  In this paper we
assume that the wire is long, so that the parameters of the system, such
as the confining potential, Fermi energy, and exchange constant $J$, do
not change significantly at the interparticle distance.  It would be
interesting to generalize our approach to the case of shorter wires and
see whether the physics discussed in this paper may be responsible for the
0.7 structure in quantum point contacts.

To test the relevance of our theory to the
experiments\cite{bkane,thomas2,reilly1,reilly2} showing plateaus at
$e^2/h$, one can check whether the experimental temperature exceeds the
exchange energy $J$.  Due to the strong exponential dependence
(\ref{eq:J}) of $J$ on the density, the uncertainly of $n$ may make the
estimation of $J$ difficult.  Instead one may be able to determine $J$
experimentally by applying magnetic field.  Indeed, if the magnetic field
$B$ exceeds a certain critical value $B_c\propto J$, the spin chain
becomes completely spin polarized.  The magnitude of the critical field
$B_c$ can be found by considering the spin-polarized state in a strong
field $B$ with a single spin-flip excitation.  The energy of such an
excitation $|g|\mu_B B$ is reduced by $2J$ due to coupling to neighboring
spins.  (Here $g$ is the Lande factor, and $\mu_B$ is Bohr magneton.) Thus
the complete polarization occurs at $B>B_c$, where
\begin{equation}
  \label{eq:B_c}
  B_c = \frac{2J}{|g|\mu_B}.
\end{equation}
By measuring the critical field $B_c$ required to achieve complete
polarization of the spin chain one can determine the exchange constant
$J$.

In the case of non-interacting electrons at zero temperature the
conductance does not depend on the magnetic field and remains $2e^2/h$
until the electron gas becomes completely spin polarized at
$B>B_c^{(0)}=E_F/4|g|\mu_B$.  In a polarizing field only one spin channel
is allowed in the wire, and the conductance reduces to $e^2/h$.  In the
case of a quantum wire at low electron density this behavior is preserved,
but the step in conductance occurs at a much lower critical field
(\ref{eq:B_c}).  Indeed, although in the presence of magnetic field
spinons are no longer the elementary excitations of a spin chain, at
$B<B_c$ one can introduce modified elementary excitations with similar
properties.\cite{psinon} Then by repeating the arguments of
Sec.~\ref{sec:isotropic} one concludes that the low energy excitations
present in the system at $T\to0$ cross the wire elastically, resulting in
$R_\sigma=0$ and total conductance $G=2e^2/h$.  At $B>B_c$ the wire is
completely spin polarized, and the spin excitations in the leads are
reflected by the wire.  This situation is completely analogous to the case
of high temperature considered in Sec.~\ref{sec:isotropic}.  In
particular, the resistance $R_\sigma$ can be found by bosonizing the
electron system in the leads and imposing the boundary condition
(\ref{eq:boundary_condition_phi_sigma}) on the field $\phi_\sigma$.  This
again leads to $R_\sigma=h/2e^2$ and reduces the conductance of the device
to $e^2/h$.  Thus one can find the critical field (\ref{eq:B_c}) and the
exchange constant $J$ by measuring the magnetic field at which the
conductance drops from $2e^2/h$ to $e^2/h$.

Apart from the experiments with GaAs quantum wires, quantization of
conductance at $G=e^2/h$ in the absence of magnetic has been observed in
carbon nanotubes.\cite{deheer} This anomalous quantization occurs when the
current is forced to flow through the narrow tip of the tube.  At small
radius of the nanotube the Coulomb interactions between electrons become
effectively stronger, and could conceivably suppress the exchange coupling
$J$ of the electron spins below the temperature.  Our result
(\ref{eq:G-high-T}) would then explain the experimental data.\cite{deheer}

\begin{acknowledgments}
  The author acknowledges helpful discussions with A.~V.  Andreev, A. M.
  Finkel'stein, L.~I. Glazman, A.~I.  Larkin, R. de Picciotto, M.
  Pustilnik, and P.~B.  Wiegmann and the hospitality of Bell Laboratories,
  where part of this work was carried out.  This work was supported by the
  U.S. DOE, Office of Science, under Contract No. W-31-109-ENG-38, by the
  Packard Foundation and by NSF Grant DMR-0214149.
\end{acknowledgments}

\appendix

\section{Estimate of the exchange constant $J$}
\label{sec:WKB}

Here we estimate the exchange constant in an infinite 1D Wigner crystal
with the lattice constant $b=1/n$.  Following the idea of
H\"ausler\cite{hausler} we evaluate $J$ for two spins at neighboring sites
$l=0$ and $l=1$ using an approximation where the only dynamical variable
is the distance $x=x_1-x_0$ between the two electrons.  In this
approximation, $x_0+x_1=b$ and all the other electrons ($l\neq0,1$) are at
fixed positions $x_l=lb$.  Then the Coulomb potential takes the form
\begin{equation}
  \label{eq:potential}
  U(x) = \frac{e^2}{\varepsilon|x|} 
               + \sum_{l\neq0,1}\!\!
               \left(\frac{e^2}{\varepsilon\big|\frac{b-x}{2}-lb\big|} 
                    +\frac{e^2}{\varepsilon\big|\frac{b+x}{2}-lb\big|}
                                \right)\!.
\end{equation}
This potential has two degenerate minima at $x=\pm b$ corresponding to
$x_0=0$, $x_1=b$ and $x_0=b$, $x_1=0$.  In the limit of strong Coulomb
potential the tunneling between these two states gives rise to
exponentially small splitting $\Delta E$ of the doublet.

The ground state wavefunction of this system is an even function of $x$,
and is therefore symmetric with respect to permutation $x_0\leftrightarrow
x_1$, while the first excited state is antisymmetric.  The two states
correspond to the values of the total spin of the two-electron system
$S=0$ and $S=1$, respectively.  Thus the energy of the two components of
the doublet can be written in terms of the electron spin operators at the
two sites as $E_0 + J\, {\bm S}_{0}\cdot{\bm S}_{1}$, where $J$ is
identified with the level splitting $\Delta E$.

Strictly speaking the infinite series in Eq.~(\ref{eq:potential})
diverges.  This is due the long range nature of the Coulomb interactions.
In practice the interactions are screened at large distances by remote
gates.  Instead of modifying the Coulomb potential to account for the
gate, it will be more convenient to simply subtract from
Eq.~(\ref{eq:potential}) a divergent constant $U(b)$.  Then the series
converges, and in the important region $|x|<3b$ the potential can be
presented in analytic form as
\begin{eqnarray}
  \label{eq:potential-regularized}
  U(x)&=&\frac{e^2}{\varepsilon b}[F(x/b)-F(1)],
  \\
  F(z)&=&\frac{1}{|z|} 
              - 2\psi\left(\frac{3-z}{2}\right)
              - 2\psi\left(\frac{3+z}{2}\right),
  \label{eq:F}
\end{eqnarray}
where $\psi(z)$ is the digamma function.

Evaluation of the energy level splitting $\Delta E$ for a particle of mass
$m$ in a double-well potential $U(x)$ is a well-known problem of quantum
mechanics,\cite{landau} and the result\footnote{The standard
  solution\cite{landau} applies the WKB approximation to the ground states
  of the oscillatory potentials near the minima of $U(x)$.  As a result it
  underestimates the level splitting (\ref{eq:level_splitting}) by a
  factor\cite{furry} of $\sqrt{\pi/e}\approx 1.075$.} is given by
\begin{equation}
  \label{eq:level_splitting}
  \Delta E = \frac{\hbar\omega}{\sqrt{e\pi}} 
             \exp\left(-\frac{1}{\hbar}\int_{-a}^{a}
                        \sqrt{2m[U(x)-\hbar\omega/2]}\,dx\right).
\end{equation}
Here $\omega=\sqrt{U''(b)/m}$ is the frequency of small oscillations near
the minima $x=\pm b$ of the potential $U(x)$, and $x=\pm a$ are the
classical turning points at energy $\hbar\omega/2$, i.e.,
$a=b-\sqrt{\hbar/m\omega}$.

To evaluate $\Delta E$ with the correct prefactor, one has to carefully
account for the small energy $\hbar\omega/2$ in the exponential.  The
resulting level splitting can be written as
\begin{equation}
  \label{eq:level_splitting_result}
  \Delta E = \frac{2}{\sqrt\pi} \sqrt{\hbar\omega^3 mb^2}\, e^\xi e^{-S_0},
\end{equation}
where
\begin{eqnarray}
  \label{eq:S_0}
  S_0 &=& \frac{1}{\hbar}\int_{-b}^{b}\sqrt{2mU(x)}\, dx,
  \\
  \xi &=& \int_0^1\left(
                       \sqrt{\frac{U''(b)b^2}{2U(bz)}} -\frac{1}{1-z}
                  \right)dz.
  \label{eq:xi}
\end{eqnarray}
An alternative solution\cite{coleman} of the problem using the instanton
technique leads to a result that can be also presented in the form
(\ref{eq:level_splitting_result})--(\ref{eq:xi}).

In order to apply this result to the evaluation of the exchange constant
$J$, one has to keep in mind that $x$ is the relative position of two
neighboring electrons, $x=x_0-x_1$, and replace the mass in
Eqs.~(\ref{eq:level_splitting_result}), (\ref{eq:S_0}) with the reduced
mass $m/2$.  One then finds $S_0=\eta\sqrt{b/a_B}$ with the numerical
coefficient
\begin{equation}
  \label{eq:eta}
  \eta=\int_{-1}^{1} \sqrt{F(z)-F(1)}\, dz \approx 2.817.
\end{equation}
Substitution of this result into Eq.~(\ref{eq:level_splitting_result})
gives the leading exponential behavior of Eq.~(\ref{eq:J}). 

The numerical parameter $\xi$ defined by Eq.~(\ref{eq:xi}) depends only on
the shape of the barrier separating the two minima of potential $U(x)$.
For the potential (\ref{eq:potential-regularized}), we find
\begin{equation}
  \label{eq:xi_value}
  \xi=\int_0^1\left(
                  \sqrt{\frac{F''(1)}{2[F(z)-F(1)]}} -\frac{1}{1-z}
                  \right)dz \approx -0.423.
\end{equation}
Substituting this result in Eq.~(\ref{eq:level_splitting_result}) we find 
the exchange constant
\begin{equation}
  \label{eq:J_result}
  J=\frac{\kappa \hbar^2}{m\sqrt[4]{b^{5}a_B^{3}}}
     \exp \left(-\eta\sqrt{\frac{b}{a_B}}\right),
\end{equation}
with $\kappa\approx 2.203$.  Expressing the prefactor in terms of the
Fermi energy, we obtain Eq.~(\ref{eq:prefactor}). 

It is worth mentioning that because of the singularity of the potential
$U(x)$ at $x=0$, the validity of the WKB approximation used in the
derivation\cite{landau} of formula (\ref{eq:level_splitting}) is limited
to $|x|\gg a_B$.  Moreover, since the potential
(\ref{eq:potential-regularized}) is not integrable up to the singularity,
it represents an impenetrable barrier.\cite{andrews} Thus the true value
of the level splitting for potential (\ref{eq:potential-regularized}) is
$\Delta E=0$.  On the other hand, the electrons in a quantum wire are not
strictly one-dimensional due to the finite width $w$ of the wire.  As a
result the singularity of the Coulomb interaction potential is cut off at
short distances $x\sim w$.  In GaAs devices $w\gtrsim a_B$, which
justifies the WKB approximation.  In carbon nanotubes it may possible to
achieve the regime $w\ll a_B$; a more sophisticated approach to the
calculation of the exchange constant $J$ is required in this
case.\cite{fogler}

\section{Resistance of a quantum wire}
\label{sec:quantum_resistance}

Let us derive the resistance (\ref{eq:R_rho_renormalized}) of an infinite
quantum wire in the regime of applied current.  The wire is modeled by the
Hamiltonian (\ref{eq:H_rho}) with the time-dependent boundary condition
(\ref{eq:boundary_condition_phi_rho}).  It is convenient to apply to the
Hamiltonian a unitary transformation
\begin{equation}
  \label{eq:U}
  U=\exp\left(-i\frac{\pi q(t)}{\sqrt2}
              \int_{-\infty}^\infty \Pi_\rho(x)dx\right),
\end{equation}
which shifts the charge field $\phi_\rho(x)\to
\phi_\rho(x)+\frac{\pi}{\sqrt2}\, q(t)$.  As a result the boundary
condition (\ref{eq:boundary_condition_phi_rho}) is replaced with
$\phi_\rho(0,t)=0$, but the Hamiltonian (\ref{eq:H_rho}) acquires a
time-dependent perturbation 
\begin{equation}
  \label{eq:perturbation}
  V = -i\hbar U^\dagger\partial_t U 
    = -\frac{\pi\hbar}{\sqrt2}\, \dot q(t) 
       \int_{-\infty}^\infty \Pi_\rho(x)dx.
\end{equation}

The perturbation (\ref{eq:perturbation}) leads to excitation of plasmons
and to dissipation of energy into the Luttinger liquid.  To find the
energy $W$ dissipated in unit time, it is convenient to diagonalize
$H_\rho$ by introducing the plasmon destruction operators
\begin{equation}
  \label{eq:b_k}
   b_k=\int\theta(kx)\sin kx
      \left(\frac{1}{\pi}\sqrt{\frac{|k|}{K_\rho}}\,\phi_\rho(x)
            +i\sqrt{\frac{K_\rho}{|k|}}\,\Pi_\rho(x)
      \right)dx,
\end{equation}
where $\theta(y)$ is the unit step function.  Note that in order to
satisfy the boundary condition $\phi_\rho(0,t)=0$ the wavefunctions of the
plasmons were chosen in the form $\varphi_k(x) = \sqrt{2/\pi}\, \theta(kx)
\sin kx$; positive and negative $k$ correspond to excitations to the right
and left of the boundary $x=0$, respectively.

Upon the transformation to the new variables (\ref{eq:b_k}), the two terms
in the Hamiltonian take the form
\begin{equation}
  \label{eq:H_transformed}
  H_\rho=\int_{-\infty}^\infty\!
           \hbar\omega_k b_k^\dagger b_k dk,
  \quad
  V=\frac{i\hbar I_0 \cos\omega t}{e\sqrt{2K_\rho}}\int_{-\infty}^\infty\!
         \frac{b_k-b_k^\dagger}{\sqrt{|k|}}\,dk. 
\end{equation}
The perturbation $V$ leads to both emission and absorption of plasmons
with energy $\hbar\omega$.  The total energy dissipated in unit time can
be evaluated using the Fermi golden rule as
\begin{equation}
  \label{eq:golden_rule}
  W = \frac{2\pi}{\hbar} 
      \left(\frac{\hbar I_0}{2e\sqrt{2K_\rho}}\right)^2
      \frac{2}{\hbar\omega}
      [(1+f_k)\hbar\omega-f_k\hbar\omega].
\end{equation}
Regardless of the values of the plasmon occupation numbers $f_k$,
expression (\ref{eq:golden_rule}) reduces to $W=\frac12 I_0^2 R_\rho$ with
the resistance (\ref{eq:R_rho_renormalized}).

\section{Dissipation of energy by a scatterer in a Fermi gas}
\label{sec:dissipation_XY}

Let us consider the dissipation of energy in a non-interacting Fermi gas
in the presence of a moving scatterer.  We assume that the single-particle
Hamiltonian has the general form
\begin{equation}
  \label{eq:single-particle-H}
  H(y,p,t) = H_0(y+q_0\sin\omega t, p).
\end{equation}
Here the hermitian operator $H_0(y,p)$ is independent of the coordinate
$y$ in the regions corresponding to the leads, $y\to\pm \infty$.  The
Hamiltonian (\ref{eq:XY}) obviously satisfies these conditions for
smoothly varying $J[y]$ after the discrete site number $l$ is replaced by
a continuous coordinate $y$.  The $y$-dependent central part of the
Hamiltonian $H_0(y,p)$ can be viewed as a scatterer with energy-dependent
reflection coefficient $\mathcal{R}(\epsilon)$.  Condition
(\ref{eq:single-particle-H}) implies that the position of the scatterer
oscillates with amplitude $q_0$.

In the limit of small $q_0$ one can expand
Eq.~(\ref{eq:single-particle-H}) and present the Hamiltonian as
\begin{equation}
  \label{eq:Taylor}
  H(y,p,t) = H_0(y,p) + q_0\sin\omega t\, \partial_y H_0(y,p).
\end{equation}
The time-dependent perturbation leads to absorption and emission of energy
quanta $\hbar\omega$ by the fermions.  The rates of these processes may be
found using the Fermi golden rule, and one obtains the increase $W$ of the
energy of the system in unit time in the form
\begin{eqnarray}
  \label{eq:Golden-W}
  W&=&\frac{2\pi}{\hbar}\hbar\omega\int\!\!\int dk dk'
      \left|\frac{q_0}{2}[\partial_y H_0]_{kk'}\right|^2 
  \nonumber\\
    &&\times[f(\epsilon_k)-f(\epsilon_k')] 
             \delta(\epsilon_k - \epsilon_{k'}+\hbar\omega). 
\end{eqnarray}
Here $k$ labels the eigenstates of Hamiltonian $H_0$ with energies
$\epsilon_k$.  The occupation numbers of these states are given by the
Fermi function $f(\epsilon_k)$.  The eigenfunctions have scattering wave
asymptotics 
\begin{equation}
  \label{eq:scattering_waves_positive}
    \psi_k(y)=\frac{1}{\sqrt{2\pi}}\times
      \left\{
      \begin{array}{ll}
         e^{iky} +r_k e^{-iky},   & \mbox{at $y\to - \infty$,}\\[1ex]
         t_k e^{iky},             & \mbox{at $y\to +\infty$.}
      \end{array}
       \right.
\end{equation}
for positive $k$ and
\begin{equation}
  \label{eq:scattering_waves_negative}
    \psi_k(y)=\frac{1}{\sqrt{2\pi}}\times
      \left\{
      \begin{array}{ll}
         t_k e^{iky},             & \mbox{at $y\to -\infty$,}\\[1ex]
         e^{iky} +r_k e^{-iky},   & \mbox{at $y\to + \infty$.}
      \end{array}
       \right.
\end{equation}
for negative $k$.  Here $r_k$ and $t_k$ are the reflection and
transmission amplitudes; the reflection coefficient is defined as
$\mathcal{R}(\epsilon_k)=|r_k|^2$.

In the limit of low frequency $\omega\to0$ expression (\ref{eq:Golden-W})
can be further simplified,
\begin{equation}
  \label{eq:W-simplified}
  W=\frac{\pi(\omega q_0)^2}{\hbar v_F^2}
    \int d\epsilon_k [-f'(\epsilon_k)]
         [\zeta_+(\epsilon_k)+\zeta_-(\epsilon_k)].
\end{equation}
Here we have approximated the energies near the Fermi level as
$\epsilon_k=\hbar v_F (|k|-k_F)$, accounted for the double degeneracy of
the energy levels $\epsilon_k$, and introduced
\begin{equation}
  \label{eq:zeta}
  \zeta_\pm(\epsilon_k)=\lim_{k'\to \pm k}|[\partial_y H_0]_{kk'}|^2.
\end{equation}
The matrix element $[\partial_y H_0]_{k,k'}$ is defined as
\begin{equation}
  \label{eq:matrix_element_definition}
  [\partial_y H_0]_{kk'}=\int dy\, 
                          \psi_{k'}^*(y)[\partial_y H_0]\psi_k(y).
\end{equation}
Integrating by parts and taking advantage of the fact that $\psi_k$ is an
eigenfunction of $H_0$, we find
\begin{equation}
  \label{eq:matrix_element}
  [\partial_y H_0]_{kk'}=(\epsilon_k-\epsilon_{k'})\int dy\, 
                          \psi_{k'}^*(y)\partial_y\psi_k(y).
\end{equation}
To evaluate $\zeta_\pm(\epsilon_k)$ we need to find the divergent at
$k'\to\pm k$ part of the integral in Eq.~(\ref{eq:matrix_element}).  Since
the divergences originate at $y\to\pm\infty$, one can use the scattering
wave asymptotics (\ref{eq:scattering_waves_positive}) and
(\ref{eq:scattering_waves_negative}) in Eq.~(\ref{eq:matrix_element}).
This results in
\begin{eqnarray}
  \label{eq:zetas}
  \zeta_+(\epsilon_k)&=&\frac{1}{\pi^2}(\hbar v_F k_F)^2 
                        [\mathcal{R}(\epsilon_k)]^2,
  \\
  \zeta_-(\epsilon_k)&=&\frac{1}{\pi^2}(\hbar v_F k_F)^2 
                        \mathcal{R}(\epsilon_k)[1-\mathcal{R}(\epsilon_k)].
\end{eqnarray}
Substituting these results into Eq.~(\ref{eq:W-simplified}), we find
\begin{equation}
  \label{eq:W-result}
  W=\frac{\hbar}{\pi}(\omega q_0)^2 k_F^2
    \int d\epsilon [-f'(\epsilon)]\mathcal{R}(\epsilon).
\end{equation}
To apply this result to the evaluation of the spin contribution to the
resistance within the XY model approximation, one should substitute
$\omega q_0=I_0/e$ and $k_F=\pi/2$.  Then Eq.~(\ref{eq:W-result}) takes
the form
\begin{equation}
  \label{eq:XY-results}
  W=\frac12 I_0^2 R_\sigma^{XY}, 
    \quad
    R_\sigma^{XY}=\frac{h}{4e^2}
                  \int d\epsilon [-f'(\epsilon)]\mathcal{R}(\epsilon).
\end{equation}
The result for the resistance coincides with Eq.~(\ref{eq:R_XY}) for the
appropriate reflection coefficient $\mathcal{R}(\epsilon) =
\theta(|\epsilon|-J)$.

\end{document}